\documentclass[12pt,preprint]{aastex}
\newcommand{\nraoblurb}{The National Radio Astronomy Observatory is
a facility of the National Science Foundation operated under cooperative
agreement by Associated Universities, Inc.}

\newcommand{\kpc}{$\,{\rm kpc}$}
\newcommand{\K}{\ensuremath{\,{\rm K}}}

\newcommand{\kms}{\ensuremath{\,{\rm km\,s}^{-1}}}

\newcommand{\hi}{{\rm H\,}{{\sc i}}}
\newcommand{\hii}{{\rm H\,}{{\sc ii}}}

\newcommand{\co}{\ensuremath{^{12}{\rm CO}}}
\newcommand{\cor}{\ensuremath{^{13}{\rm CO}}}

\newcommand{\hisa}{{\rm H\,}{{\sc i}{\rm~SA}}}
\newcommand{\hiea}{{\rm H\,}{{\sc i}{\rm~E/A}}}
\newcommand{\titleI}{\thinspace \bf {\small I}}
\newcommand{\titleII}{\thinspace \bf {\small II}}

\bibpunct[; ]{(}{)}{;}{a}{}{; }
\slugcomment{draft \# 5 --- lda, \today}

\shorttitle{Distances for H\titleII\ Regions}
\shortauthors{Anderson et al.}
\begin{document}

\title{Resolution of the Distance Ambiguity for Galactic H\titleII\ Regions}

\author{L. D. Anderson\altaffilmark{1} \& T. M. Bania\altaffilmark{1}}

\altaffiltext{1}{Institute for Astrophysical Research,
725 Commonwealth Ave., Boston University, Boston MA 02215, USA.}

\begin{abstract}
We resolve the kinematic distance ambiguity for 266 inner Galaxy \hii\
regions out of a sample of 291 using existing \hi\ and \cor\ sky
surveys.  Our sample contains all \hii\ regions with measured radio
recombination line (RRL) emission over the extent of the \cor\ Boston
University---Five College Radio Astronomy Observatory Galactic Ring
Survey ($18 \arcdeg\ < l < 55 \arcdeg$ and $|b| < 1$) and contains
ultra compact, compact, and diffuse \hii\ regions.  We use two methods
for resolving the distance ambiguity for each \hii\ region: \hi\
emission/absorption (\hiea) and \hi\ self-absorption (\hisa).  We find
that the \hiea\ and \hisa\ methods can resolve the distance ambiguity
for 72\% and 87\% of our sample, respectively. When projected onto the
Galactic plane, this large sample appears to reveal aspects of
Galactic structure, with spiral arm-like features at Galactocentric
radii of 4.5 and 6 kpc, and a lack of \hii\ regions within 3.5 kpc of
the Galactic center.  Our \hii\ regions are approximately in the ratio
of 2 to 1 for far verses near distances.  The ratio of far to near
distances for ultra-compact \hii\ regions is 2.2 to 1.  Compact \hii\
regions are preferentially at the near distance; their ratio of far to
near distances is 1.6 to 1.  Diffuse \hii\ regions are preferentially
at the far distance; their ratio of far to near distances is 3.8 to 1.
This implies that the distinction between ultra compact and compact
\hii\ regions is due largely to distance, and that the large angular
size of diffuse \hii\ regions is not due solely to proximity to the
Sun.
\end{abstract}


\section{INTRODUCTION\label{sec:intro}}
\hii\ regions are the clearest tracers of massive star formation
because of their visibility across the Galactic disk at cm-wavelengths.
Without a known distance, however, the physical properties of \hii\
regions remain unknown.  Distances allow one to transform
measured properties such as flux and angular size into physical
properties such as luminosity and physical size.  The knowledge of these
properties for \hii\ regions is vital to understanding the process of
star formation in our Galaxy.

The simplest and most common way to estimate the distance to an \hii\
region with a measured velocity is to assume a Galactic rotation
curve.  Rotation curves generally model axially symmetric circular
orbits such that a source's velocity is a function only of
Galactocentric distance.
There have been many models of Galactic rotation proposed using
various tracers \citep[e.g][]{burtongordon78, clemens85, brand86,
fich89, mcclure07}.  All rotation curves have some amount of intrinsic
error because they are simplified models of a more complicated
rotation pattern.  Additionally, when transforming velocities into
distances, velocities that depart from circular rotation (due to
streaming motions) will cause errors in the derived distance.
Streaming motions result in velocity deviations of $\sim 10$ \kms\
\citep{burton71, stark89}, which cause a distance uncertainty up to 2
kpc over the longitude range of our present study.

While the procedure of transforming velocities into distances is
straightforward in the outer Galaxy, most distances in the inner Galaxy
are degenerate: for each velocity there are two possible distances.
These two distances (a ``near'' and a ``far'' distance) are spaced
equally about the tangent point distance.  This problem has become known
as the kinematic distance ambiguity (KDA).  For only one positive
velocity along each line of sight, the tangent point velocity, is the
degeneracy not present.  Determining whether the \hii\ regions in our
sample lie at the near or the far distance is the goal of this paper.

\begin{figure}
\plotone{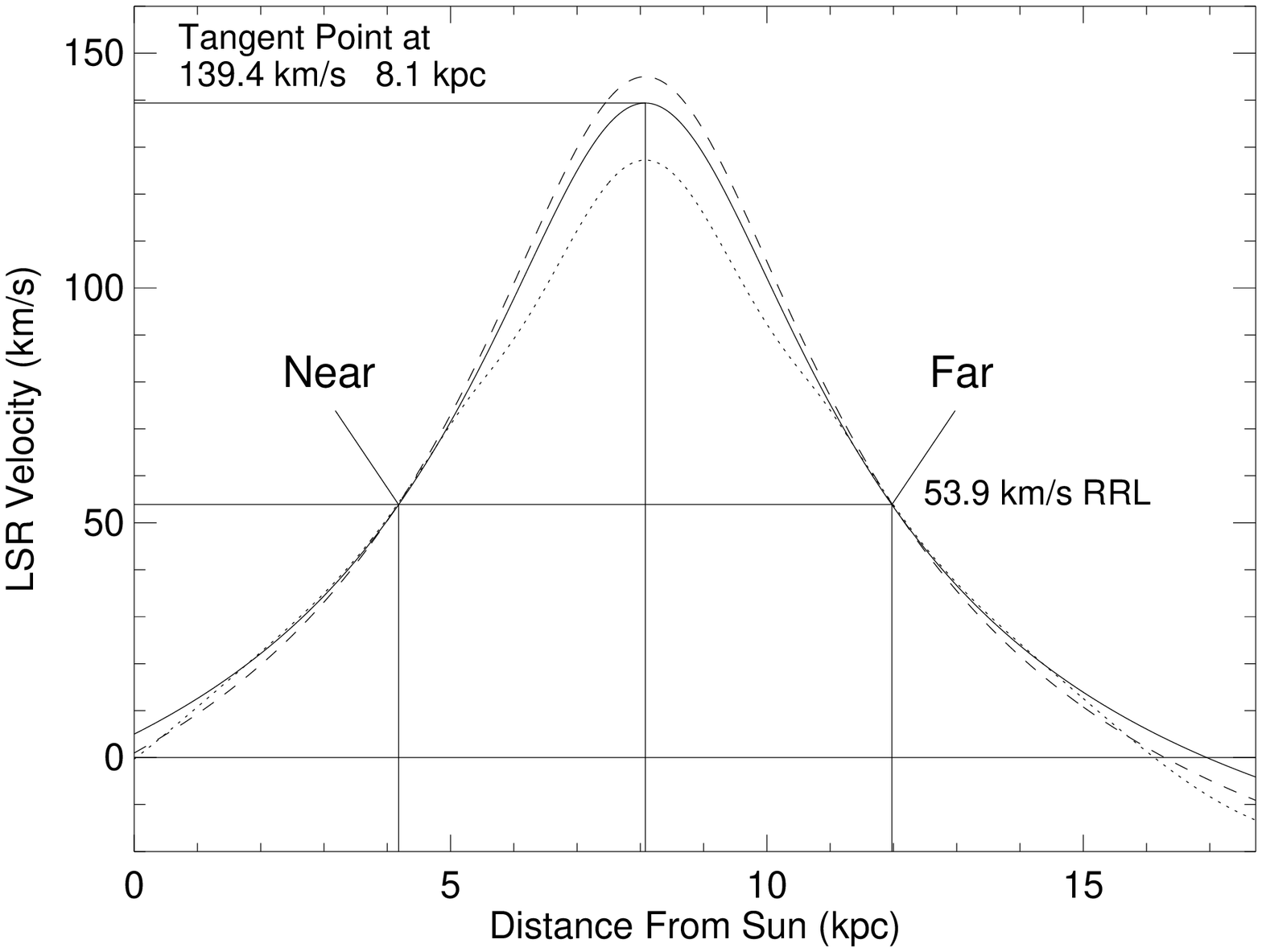}

\caption{The inner Galaxy Kinematic Distance Ambiguity.  Shown is the
LSR velocity plotted as a function of distance from the Sun for the line
of sight toward the \hii\ region G18.15$-$0.28, which has an observed
radio recombination line (RRL) velocity of +53.9\kms\ \citet{lockman89}.  Three different
models for Galactic rotation, \citet[][dotted]{clemens85},
\citet[][dashed]{brand86}, and \citet[][solid]{mcclure07}, give
identical distances for both the near and far kinematic distances.
G18.15$-$0.28 must be located at the 4.2 kpc near distance because \hi\
absorption is not seen at velocities between the RRL and the tangent
point terminal velocity of +139.4\kms\ (see \S \ref{sec:hiea_protocol}).}

\label{fig:velcurve}
\end{figure}

One method that has proved effective at resolving the KDA towards \hii\
regions is \hi\ Emission/Absorption, hereafter \hiea.  All \hii\ regions
emit broadband thermal free-free continuum radiation in the
cm-wavelength regime.  Neutral \hi\ gas between the \hii\ region and the
observer will absorb the thermal continuum if the brightness temperature
of the \hi\ is less than that of the \hii\ region at 21 cm.  Because the
continuum emission is broadband, and not limited to a particular
frequency (velocity), all foreground \hi\ has the potential to absorb
the \hii\ region's continuum.  In the first quadrant, the LSR velocity
increases with distance from the Sun, reaches a peak at the tangent
point, and decreases with distance thereafter (see Figure
\ref{fig:velcurve}).  Thus \hii\ regions at the near distance will show
\hi\ absorption lines from foreground clouds at velocities up to their radio
recombination line (RRL) velocity.  If an \hii\ region is at the far
distance, it will show absorption from foreground \hi\ clouds with
velocities up to the tangent point velocity.  The \hiea\ method
therefore relies on \hi\ absorption at velocities between the RRL and
tangent point velocity to distinguish between the near and the far
distance.

Many authors have used \hiea\ to resolve the KDA for \hii\
regions. \citet{kuchar94} used both pointed Arecibo 21 cm \hi\
observations (HPBW of $4\arcmin$) and also Boston University---Arecibo
Observatory Galactic \hi\ Survey data \citep[hereafter BUAO \hi\ survey;
][]{kuchar93} to resolve the KDA for 70 Galactic \hii\ regions.
\citet{fish03} resolved the KDA using the \hiea\ method for a sample of
20 UC \hii\ regions using the VLA in A- or BnA-configurations
($1\arcsec$ to $6\arcsec$ synthesized beam).  \citet{kolpak03}
measured the 21 cm \hi\ absorption spectrum towards 49 \hii\ regions
using the VLA in C-configuration ($16 \arcsec$ synthesized beam).  The
high success rate of these efforts proves the utility of \hiea\
experiments in \hii\ region distance determinations.  

A second method that has been used to resolve the KDA is \hi\
self-absorption, hereafter \hisa.  \hisa\ is useful for resolving the
KDA for molecular clouds.  Cold foreground \hi\ will absorb against
warmer background \hi\ at the same velocity.  \citet{liszt81}
hypothesized that Galactic molecular clouds must contain residual \hi, a
result that has been confirmed by many observations
\citep[e.g.][]{wannier91, kuchar93, williams96}.  The \hi\ gas
associated with molecular clouds is cold ($\sim 10 \K$) compared the the
warm \hi\ in the ISM ($\sim 100 \K$).  The \hi\ inside a molecular cloud
at the near distance will absorb against the warm background \hi\ at the
same LSR velocity that lies at the far distance.  The \hi\ inside a
molecular cloud at the far distance shows no such absorption as there is
no background \hi\ at the same velocity.  Thus the signature of a cloud
at the near distance is molecular emission at the same velocity and with
the same line width as an \hi\ absorption feature.  Since \hii\ regions
are almost always associated with molecular gas (Anderson et al. 2008;
hereafter Paper I), the distance to \hii\ regions can be found using
\hisa.

Molecular clouds frequently display \hisa\ features
\citep[e.g.][]{knapp74}.  A theoretical study by \citet{flynn04} showed
that all model molecular clouds could produce \hisa. \citet{jackson02}
found a strong \hisa\ signal associated with the molecular cloud located
at ($l,b$)=($45\fdg6,+0\fdg3$) using the BUAO \hi\ survey \citep{kuchar93} and
the BU--FCRAO \cor\, Galactic Ring Survey data\footnote[1]{Data
available at http://www.bu.edu/galacticring/} \citep[GRS;
][]{jackson06}.  \citet{goldsmith05} found that the \cor\ molecule is a
very good tracer of \hisa.
More recently, \citet{busfield06} used \hisa\ analysis to determine the
distances towards massive young stellar object candidates.

One can also use formaldehyde (${\rm H_2CO}$) absorption to resolve the
KDA.  This method is identical to the \hiea\ method, except that the
absorber is ${\rm H_2CO}$ instead of \hi.  Again, it is the broadband
nature of the radio continuum emission of \hii\ regions that makes this
method possible, as the radio continuum and ${\rm H_2CO}$ lines are both
bright in the cm-regime.  The ${\rm H_2CO}$ absorption method was used
by \citet{wilson72} in a study of 73 discrete radio continuum peaks in
the Galactic plane, 49 of which are \hii\ regions.  \citet{downes80}
used this method on a larger sample of Galactic radio sources.  More
recently, \citet{araya02}, \citet{watson03}, and \citet{sewilo04} used
${\rm H_2CO}$ absorption towards ultra compact (UC) \hii\ regions to
resolve the KDA towards a combined 108 out of 147 sources.



Because of the high filling factor of \hi\ compared to ${\rm H_2CO}$,
\hi\ absorption methods are more robust than ${\rm H_2CO}$ absorption
methods.  In ${\rm H_2CO}$ studies there may not be a molecular cloud
with appreciable ${\rm H_2CO}$ that has a velocity that lies between the
\hii\ region and the tangent point velocities.  In this case the near
distance will be mistakenly chosen since there will be no absorption
line between the source velocity and the tangent point velocity.  On the
basis of their observations, \citet{watson03} estimate that there is one
cloud containing measurable ${\rm H_2CO}$ every 2.9 \kpc.  For \hi\,
there is average one cloud every 0.7 kpc \citep{bania84}.  The chance of
the misassignment increases as the difference between the near distance
and the tangent point distance decreases, making the determination using
${\rm H_2CO}$ absorption for sources near the tangent point less
robust than the determination made using \hi\ absorption.


\section{DATA}

\subsection{The VLA Galactic Plane Survey}

For the \hiea\ analysis we use data from the VLA 21cm \hi\ Galactic
Plane Survey \citep[VGPS; ][]{stil06}.  The VGPS sky coverage spans $18
\arcdeg < l < 67\arcdeg$ with $|b|$ varying from $1\fdg3$ to $2\fdg6$
over the longitude range.  The angular resolution of the VGPS is
$1\arcmin$.  Short-spacing information was obtained by making an \hi\
survey with the NRAO Green Bank Telescope.  The spectral resolution of
the VGPS is 1.56 \kms, with a channel width of 0.824 \kms.  The RMS
noise in the VGPS is $\sim 2$ K per 0.824 \kms\ channel.  In addition to
the \hi\ data cubes, the VGPS also produced continuum maps from the
portions of each spectrum without \hi\ line emission.  These 21cm
continuum maps have the same $1\arcmin$ resolution as the line emission
data cubes.

\subsection{The Southern Galactic Plane Survey}

For the range $15\arcdeg < l < 18\arcdeg$ we use the Southern Galactic
Plane Survey \citep[SGPS; ][]{mcclure05}.  The SGPS sky coverage spans
$253\arcdeg < l < 358\arcdeg$ and $5\arcdeg < l < 20\arcdeg$ over the
latitude range $|b| \le 1.5$.  The SGPS combines data from the Australia
Telescope Compact Array and the Parkes Radio Telescope.  The angular
resolution of the SGPS is $\sim 2\arcmin$.  The spectral resolution of
the SGPS is 0.8 \kms.  The RMS noise in the SGPS is 1.6 K.  The
continuum data for the SGPS are not yet public over the longitude range
of interest here, so we use the SGPS data at large negative velocities
as a proxy.

\subsection{{\bf The $^{\bf 13}$}CO Galactic Ring Survey\label{GRS}}

For the \hisa\ analysis we use both the \hi\ surveys and the \cor\ GRS.
The GRS sky coverage spans 74 square degrees, from $18 \degr < l < 56
\degr$ and $\vert b\vert < 1 \degr$ with additional, incomplete sky
coverage from $14 \degr < l < 18 \degr$ over the same latitude range.
The GRS has an angular resolution of $46\,\arcsec$, a spectral
resolution of $0.21 \kms$, and $22 \, \arcsec$ angular sampling.  The
typical spectral RMS of the GRS is 0.13 K (T$_A^*$).

\subsection{The H\titleII\ Region Sample}

Our sample of 291 \hii\ regions is that of Paper I, which used the GRS
to associate molecular gas clumps with \hii\ regions.  This sample
represents all \hii\ regions with measured RRL emission over the extent
of the GRS.  The sample contains UC \hii\ regions \citep[compiled from
][]{araya02, watson03, sewilo04}, classical compact \hii\ regions
\citep[compiled from ][]{lockman89}, and diffuse regions \citep[compiled
from ][]{lockman96}.  Here we follow the naming convention of Paper I.
The first letter of the \hii\ region name (``U'', ``C'', or ``D'')
refers to its classification (UC, compact, or diffuse).  In Paper I, all
sources were confirmed to be \hii\ regions by examining their infrared
and radio continuum emission.

There are a number of UC \hii\ regions in this sample that have multiple
RRL velocities along the line of sight, indicating multiple \hii\
regions.  In Paper I we were able to determine which RRL velocity is
truly from the UC \hii\ region for 10 cases on the basis of the
molecular emission near these \hii\ regions.  We expunge the 10 \hii\
regions at the velocity without molecular emission from our sample.
While the RRL emission indicates that there are \hii\ regions at these velocities, they are
unlikely to be UC \hii\ regions; their removal results in a cleaner
sample.  For lines of sight with multiple \hii\ regions, the \hiea\
method is unable to distinguish which \hii\ region is causing the
absorption.

Table \ref{tab:1} summarizes the kinematic properties of our \hii\
regions derived using the \citet{mcclure07} rotation curve.  Listed in
Table \ref{tab:1} are the source name, the Galactic longitude and
latitude, the RRL velocity, the Galactocentric radius, the near
distance, the far distance, and the tangent point distance.  Typical
measurement errors on the RRL velocity are 2 \kms\ (see data compiled in
Paper I).  This error is insignificant compared to uncertainties in the
rotation curve model, which may be 10 \kms.  There are 25 lines of sight
with multiple RRL velocities.  The source names for these \hii\ regions
are denoted with an extra ``a'' or ``b'' (e.g. D17.25$-$0.20a).

\section{RESOLVING THE KDA \label{sec:resolve}}

We perform two independent analyses on each \hii\ region, using both the
\hiea\ and \hisa\ techniques.
There are many benefits to using both methods.  Some sources may be
better suited to one method.  The use of both methods also allows us to
test the robustness of the methods themselves.
The \hiea\ and \hisa\ methods are complementary.  A lack of strong
absorption increases the chance that a source will be assigned to the
near distance in the \hiea\ analysis, as it may not show absorption
between the RRL velocity and the tangent point velocity.  A lack of
absorption in the \hisa\ analysis, however, implies the far distance.  A
null result in the \hiea\ analysis therefore implies the near distance,
while a null result in the \hisa\ analysis implies the far distance.

\subsection{H\titleI\ Emission/Absorption \label{sec:hiea}}
The \hiea\ method studies \hi\ absorption from foreground \hi\ clouds
absorbing against the broadband continuum emission from a background
\hii\ region.
The absorption is usually, and most easily, detected in the difference
between on-- and off--source spectra as follows:
\begin{equation} \label{eqn:dt} \Delta T(v) = [T_{\rm off}(v) - T_{\rm on}(v)] = T_{\rm c}(1 - e^{-\tau (v)}), \end{equation}
where $T_{\rm on}(v)$ and $T_{\rm off}(v)$ are the on-- and off--source
\hi\ intensity at velocity $v$, $T_{\rm c}$ is the continuum brightness
temperature, and $\tau(v)$ is the optical depth of the absorbing gas at
velocity $v$.  As can be seen from Equation \ref{eqn:dt}, absorption
features will appear as positive values of $\Delta T$.


%
%

We use the \hi\ survey data to derive the $T_{\rm on}$ and $T_{\rm off}$
spectra for each source.  We select on-- and off--source regions using
the 21cm continuum maps and extract spectra from all voxels that fall
within these on-- and off--source regions.  We then average all the
on--source spectra to create an unweighted average on--source spectrum,
and do the same for the off--source spectra.

There are many factors to consider when selecting which on-- and
off--source locations will produce the strongest and most reliable
absorption signal.  We find that the following four factors are the most
important when selecting the on-- and off--source regions from which the
spectra will be extracted:
\begin{enumerate}

\item The on-- and off--source region should be located as near to each
other as possible so the spectra share the same background.

\item The on--source region should be small and localized on the
strongest continuum emission, as this will produce the strongest
absorption.  The VGPS RMS noise is $\sim 2$ \K\ channel$^{-1}$ (the RMS
noise of the SGPS is 1.6 K channel$^{-1}$), which is large when compared
to the absorption signal for weak continuum sources.  Therefore, the
on--source region must also be large enough to significantly reduce this
noise in the resultant average on--source spectrum.

\item The off--source region should entirely encircle the on--source
spectrum in an attempt to account for gradients in the background
emission.

\item The off--source region should be located off the continuum source
and away from other continuum sources so the absorption signal is not
lost when the difference spectrum is created.

\end{enumerate}

We find that satisfying criterion (1) at the expense of the other
criteria gives the most reliable results.  To some extent, these
criteria are mutually exclusive.  For example, a large diffuse source
has difficulty satisfying conditions (1) and (2) and (3) simultaneously
since the strongest continuum locations are far away from the locations
without continuum emission.  Weak continuum sources and complicated
regions require a compromise between these criteria.

We use these criteria to select on-- and off--source regions
individually for each \hii\ region.  In the 21cm continuum images,
we define the on--source region as all contiguous pixels associated with
each \hii\ region that are above a threshold evaluated individually for
each region.  We then define the off--source region as being those
pixels that both encircle the on--source pixels, and also are free from
continuum emission.

There are two sources of noise that affect the reliability of a 21\,cm
absorption line detection: receiver noise and true variations in the
emission profile.  The calculation of emission fluctuations is of course
not free from receiver noise, so these two values do not add in
quadrature.  
We calculate the RMS receiver noise for each spectrum
individually from the channels devoid of \hi\ emission.  We estimate the
true fluctuations in the \hi\ background by computing the standard
deviation of values in the off--source spectrum at each velocity:
\begin{equation}
\sigma_{T} {\rm (}v {\rm )} = \left \{ \frac{1}{N}
\displaystyle\sum_{i=0}^n [T_{{\rm off,} i}(v) - \overline{T}_{\rm off}(v)]^2
\right \}^ {1/2},
\end{equation}
where the summation is carried out over all pixels in the off--source
region and $\overline{T}_{\rm off}(v)$ is the average value in the
off--source region.  We require the difference spectrum, $\Delta T$, to
be larger than $5\sigma_{rms}$ and larger than $1\sigma_{T}$ for a
positive detection of \hi\ absorption, criteria similar to that of other
authors \citep[e.g. ][]{payne80, kuchar94}.

We find that fluctuations in the background emission are not well
accounted for by our error estimates.  \hii\ regions should have
enough neutral hydrogen just beyond the ionization front to
produce absorption.
For each source we test for absorption within 10 \kms\ of the RRL
velocity.  Because \hii\ regions expand as they age, we allow as much as
a 10 \kms\ difference between the RRL velocity and the absorption
velocity.  The KDA resolution for sources that do not show {\it any}
absorption within 10 \kms\ of the RRL velocity should be regarded with
suspicion since the absorption signal must be very weak, and may be due
to background fluctuations.

To limit misassignments to the near distance, we assign all sources
whose source velocity is within 10 \kms\ of the tangent point velocity
to the tangent point distance.  The reliability of a \hiea\ distance
assignment decreases as the source velocity approaches the tangent point
velocity.  As the two velocities approach one another, the probability
of a cold intervening cloud decreases, and therefore a near distance
assignment becomes increasingly probable and the KDA resolution
increasingly problematic.

\subsection{H\titleI\ Self-Absorption \label{sec:hisa}}

Cold \hi\ lying in front of warm background \hi\ will absorb the warmer
emission, a process known as \hi\ self-absorption.  Molecular clouds are
composed partly of neutral hydrogen.  This neutral gas is cold compared
to the warmer \hi\ gas in the interstellar medium.  \hii\ regions are
created by massive stars inside molecular clouds, and show a strong
association with molecular gas \citep[][Paper I]{churchwell90, russeil04}.
The use of the \hisa\ method for the molecular gas associated with \hii\
regions provides another, but less direct, KDA resolution technique.

There must be significant cold \hi\ gas to produce measurable absorption
against the warm \hi\ background.  The intensity of an \hisa\ features at
LSR velocity $v$ \citep[see ][]{levinson80} is
%
\begin{equation}\Delta T_{\rm SA}(v) = (T_b(v) - T_s) (1 - e^{-\tau(v)}) , \end{equation}
where $\Delta T_{\rm SA} (v)$ is the depth of the self-absorption line
at velocity $v$, $T_b(v)$ is the intensity of the warm \hi\ background
at velocity $v$, $T_s$ is the \hi\ spin temperature, and $\tau(v)$ is
the optical depth at velocity $v$.  In the average \hi\ spectra we
create for the \hiea\ analysis the $1\sigma_{rms}$ is generally $\sim
0.2\K$.  We find, however, that variations of $\sim 10 \K$ are necessary
to distinguish \hisa\ from background fluctuations.  For $T_s = 10 \K$,
and $T_b = 100 \K$, we calculate that the minimum optical depth required
to produce measurable \hisa\ is $\sim 0.12$.  To convert from line
parameters to column density of \hi\, we use the equation
\begin{equation}N{\rm(HI)} = 1.82 \times 10^{18} ~T_s ~\tau_0 ~\Delta V, \end{equation}
where $\Delta V$ is the FWHM of the absorption line, and $\tau_0$ is the
optical depth at line center.  If we assume $\Delta V = 5 \kms$, a
column density of $\sim 1 \times 10^{19}$ \hi\ atoms is required to
produce measurable \hisa.

There should be sufficient residual cold \hi\ inside most molecular
clouds to produce a measurable \hisa\ signal.  Cosmic rays maintain a
small population of cold neutral \hi\ inside all molecular clouds.
Assuming molecular clouds are comprised of $\sim 0.2\%$ H nuclei
\citep{burton78, li03}, a column density of $6 \times 10^{21}$ ${\rm
H_2}$ molecules is required to create the necessary column density of
\hi\ for \hisa.  Using the conversions of \citet{simon01}:
\begin{equation}
N({\rm H_2}) = 4.92 \times 10 ^{20}\, T_{\rm 13CO}\, \Delta v \simeq 5 \times 10^{20}\,W_{\rm 13CO},
\end{equation}
a $W_{\rm 13CO}$ value of $\sim 10 \K \kms$ is required to produce
measurable \hisa\ from cold \hi\ in a molecular cloud.  Integrated
intensity values of this magnitude are frequently found towards \hii\
regions (Paper I).

The analysis of \hisa\ in molecular clouds is complicated by \hii\
regions for two reasons: (1) \hii\ regions are strong continuum sources
and produce absorption at many velocities along the line of sight; and
(2) \hii\ regions heat molecular clouds, which decreases the
self-absorption signal.  To eliminate absorption from continuum sources,
we locate all significant continuum features in the VGPS and SGPS data
sets.  We find in the \hiea\ analysis that locations with VGPS or SGPS
21\,cm continuum intensities of $\sim 20 \K$ can produce an absorption
signal.  This 20~K limit was also chosen by \citet{gibson05} to exclude
continuum locations from their \hisa\ analysis.  In addition to discrete
continuum sources, there is diffuse continuum emission in the plane of
the Galaxy.  This diffuse emission must be subtracted off to gauge a
continuum source's intensity reliably.  We estimate the diffuse
background by filtering the continuum images with a $10 \arcmin$ median
filter.  We then subtract this background image from the original
continuum image and locate all pixels that are above $20 \K$.  We
exclude the positions of these pixels from further \hisa\ analysis.

To solve the problem of local heating by \hii\ regions, we search for
absorption signals in the larger molecular cloud associated with the
\hii\ region, and not in the molecular material immediately surrounding
the \hii\ region.  The larger clouds we use entirely surround and
encompass the molecular clumps.  We find that in general the molecular
gas immediately surrounding \hii\ regions produces poor absorption
signals, presumably due to the increased heating.  Also, for some \hii\
regions (especially UCs) the continuum and \cor\ clump are entirely
co-spatial.  Therefore, a larger region must be analyzed for these \hii\
regions since excluding all positions with a continuum intensity greater
than 20 K would exclude all emission from the associated molecular
clump.  Further, the size scale of the molecular gas clumps associated
with \hii\ regions (a few arcminutes in diameter, see Paper I) makes
\hisa\ difficult to distinguish from background fluctuations in the VGPS
and SGPS, which are also of the same size scale.  The larger associated
clouds are likely colder and often have filamentary morphologies that
make \hisa\ signals easier to identify in both spectra and integrated
intensity images.

To identify these larger molecular clouds, we create \cor\ integrated
intensity, $W_{\rm 13CO}$, images of each \hii\ region with associated
\cor\ (253 regions) using the \cor\ line center and line width from
the analysis of Paper I.  We then locate pixels in this integrated
intensity image that are contiguous with the molecular material found
in Paper I in two trials: one trial where all selected pixels have
values greater than 75\% of the peak value found in Paper I, and a
second trial where the level is set to 85\% of the peak value.  For
bright \cor\ clumps, the 75\% level selects regions that are too large
and possibly merged with other clouds.  For dim \cor\ clumps the 85\%
level selects sources that are too small.  One of these two levels
reliably selects a single molecular cloud for nearly all sources.  For
a few sources, the 75\% level selects regions that are still within
the radio continuum emission area, and therefore a lower threshold is
necessary.  We also require all pixels to have values greater than 5
times the RMS background level in the integrated intensity image.
This criterion prevents clouds with weak \cor\ emission from becoming
too large and merging with background emission.

In the \hisa\ analysis, locating off--source positions that sample the
same background and do not cause absorption is extremely difficult and
unreliable.  This difficulty primarily stems from the large angular size
of the molecular cloud that we are probing.  Our criterion (1) for the
selection of on-- and off--source regions in the \hiea\ analysis, which
also applies to the \hisa\ analysis, mandates on-- and off--pairings
that are close on the sky.  The large size of the molecular cloud makes
this criterion nearly impossible to satisfy, and hence would lead to many
false absorption signals.  Moreover, as shown in \S\ref{sec:discussion},
a \hisa\ signal may be caused by low column densities of \cor, and
therefore it is difficult to find off--positions without absorption.

\section{H\titleII\ Region Distances}
\subsection{H\titleI\,Emission/Absorption Protocol \label{sec:hiea_protocol}}
\begin{figure}
\plotone{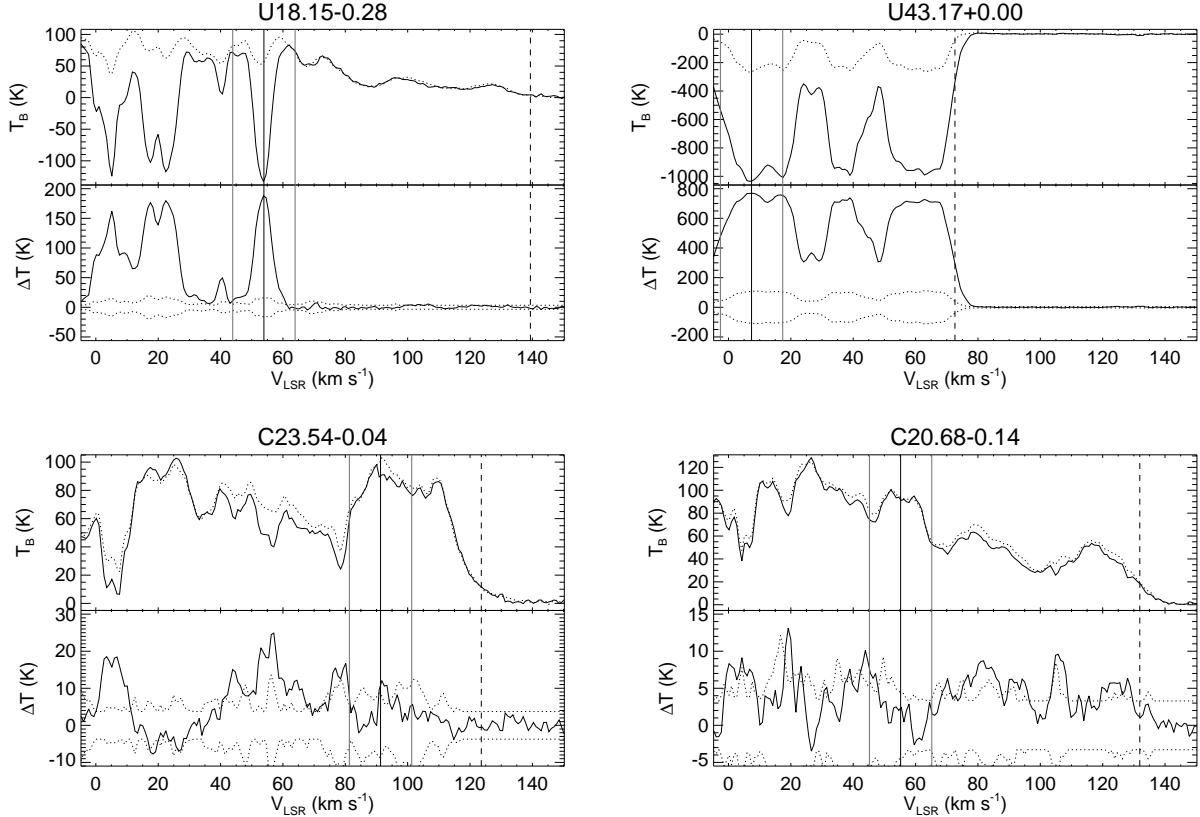}
  
\caption{Example \hiea\ analysis for four \hii\ regions.  The top panel
of each plot shows the on-- (solid line) and off-- (dotted line) source
average \hi\ spectra.  The bottom panel of each plot shows the
difference between the off-- and on--source spectra (solid line) and our
error estimates (dotted lines).  The three solid vertical lines mark the
RRL velocity and $\pm$10 \kms\ of the RRL velocity, and the dashed
vertical line marks the tangent point velocity.  Sources at the far
distance will show absorption between the RRL velocity and the tangent
point velocity.  Sources at the near distance will show absorption only
up to the RRL velocity.  Top row: example near (left) and far (right)
sources which we assign with high confidence (confidence parameter = A).
Bottom row: example near (left) and far (right) sources which we assign
with low confidence (confidence parameter = B).}

\label{fig:hiea}
\end{figure}

We determine whether the source lies at the near or the far distance
using the analysis summarized in Figure \ref{fig:hiea}, which shows
example spectra for four \hii\ regions.  The top panel in each plot shows the
on--source (solid line) and off--source (dotted line) average \hi\
spectra.  The bottom panel shows the difference spectrum ($\Delta T$).
The three vertical lines on the left mark the RRL velocity, and the $\pm
10 \kms$ area.  The vertical dashed line on the right marks the tangent
point velocity \citep{mcclure07}.  The dotted lines in the bottom panel
show our error estimates (the larger of $5\sigma_{rms}$ and
$1\sigma_{T}$), see \S\ref{sec:hiea}.

We assign a confidence parameter to each source based on the strength of
the absorption features and RRL velocity with respect to the tangent
point velocity.  This qualitative parameter indicates our confidence
that the \hiea\ KDA resolution is correct.  We are very confident in the
KDA resolutions for sources with a confidence parameter value of ``A''.
We assign sources with significant absorption features that are
generally at the same level as the noise a confidence of ``B''.  We also
assign a confidence ``B'' to sources for which the RRL velocity and the
tangent point velocity are close, as the probability of intervening gas
is low.  There are 47 sources for which we are unable to assign a
distance with any confidence due to weak or absent absorption features.

\begin{figure}
\plotone{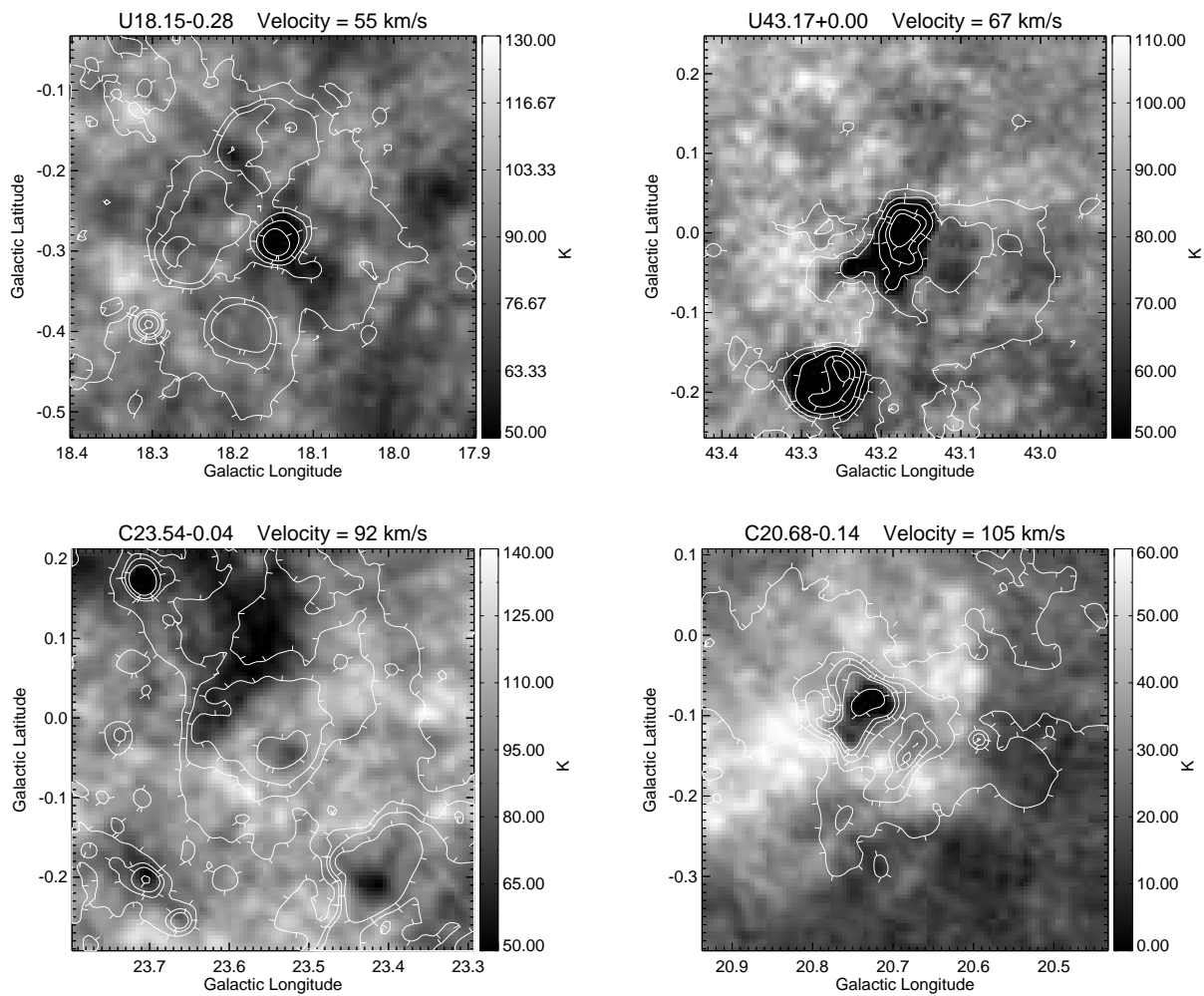}

\caption{Single channel VGPS images for the Figure \ref{fig:hiea} \hii\
regions showing the \hi\ intensity at the highest velocity with detected
absorption.  The contours are of 21cm VGPS continuum.  Tick marks on the
contours point towards decreasing values of 21cm continuum intensity.
Top row: example near (left) and far (right) sources which we assign
with high confidence (confidence parameter = A).  Bottom row: example
near (left) and far (right) sources which we assign with low confidence
(confidence parameter = B).}

\label{fig:hiea_images}
\end{figure}

Background fluctuations caused by on-- and off--source spectra drawn
from different regions can cause false absorption signals.  This is the
largest source of uncertainty in our analysis.  To separate true \hiea\
signals from background fluctuations, which may be caused by \hisa, we
require that the absorption morphology matches the continuum morphology.
We verify the \hiea\ KDA resolution for all sources by examining the
single channel 21cm \hi\ position-position intensity image at the
velocity of highest absorption with VGPS or SGPS 21cm continuum contours
overlaid.  Example images for the same sources whose spectra are shown
in Figure \ref{fig:hiea} are shown in Figure \ref{fig:hiea_images}.

\subsection{H\titleI\ Self-Absorption Protocol}
\begin{figure}
\plotone{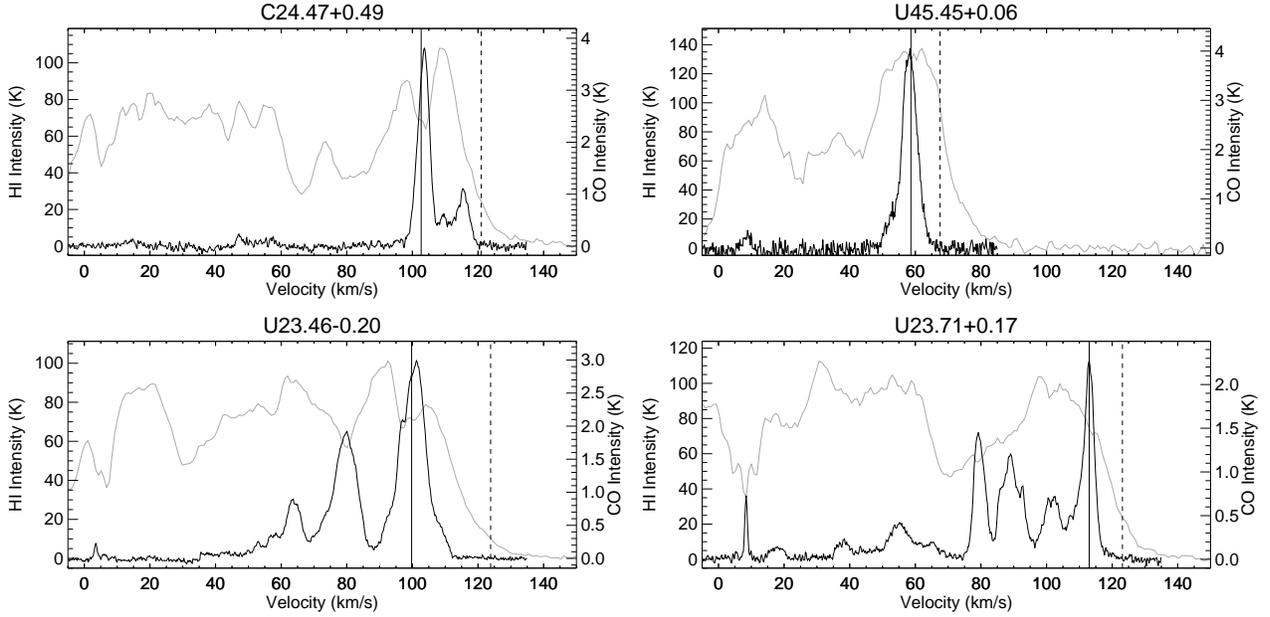}

\caption{Example \hisa\ analysis for four \hii\ regions.  The spectrum
shown in gray is \hi\ while the spectrum shown in black is \cor.  The
solid vertical line marks the velocity of the \cor\ associated with the
\hii\ region found in Paper I, and the dashed vertical line marks the
tangent point velocity.  Sources at the near distance show \hi\
absorption at the velocity of \cor\ emission.  Sources at the far
distances show no \hi\ absorption at the velocity of \cor\ emission.
Top row: example near (left) and far (right) sources which we assign
with high confidence (confidence parameter = A).  Bottom row: example
near (left) and far (right) sources which we assign with low confidence
(confidence parameter = B).}

\label{fig:hisa}
\end{figure}

We determine if the source lies at the near or the far distance using
the analysis summarized in Figure \ref{fig:hisa}, which shows examples
for four \hii\ regions.  In Figure \ref{fig:hisa}, the gray curve is the
average \hi\ spectrum and the black curve is the average \cor\ spectrum.
The vertical solid line marks the velocity of the associated \cor\ gas
found in Paper I, while the vertical dashed line shows the tangent point
velocity.  We visually examine each plot for absorption features in the
\hi\ spectrum at the velocity of the associated \cor\ emission.  These
absorption features must also share nearly the same line width in \hi\
and \cor\ to be true self-absorption features.

\begin{figure}
\plotone{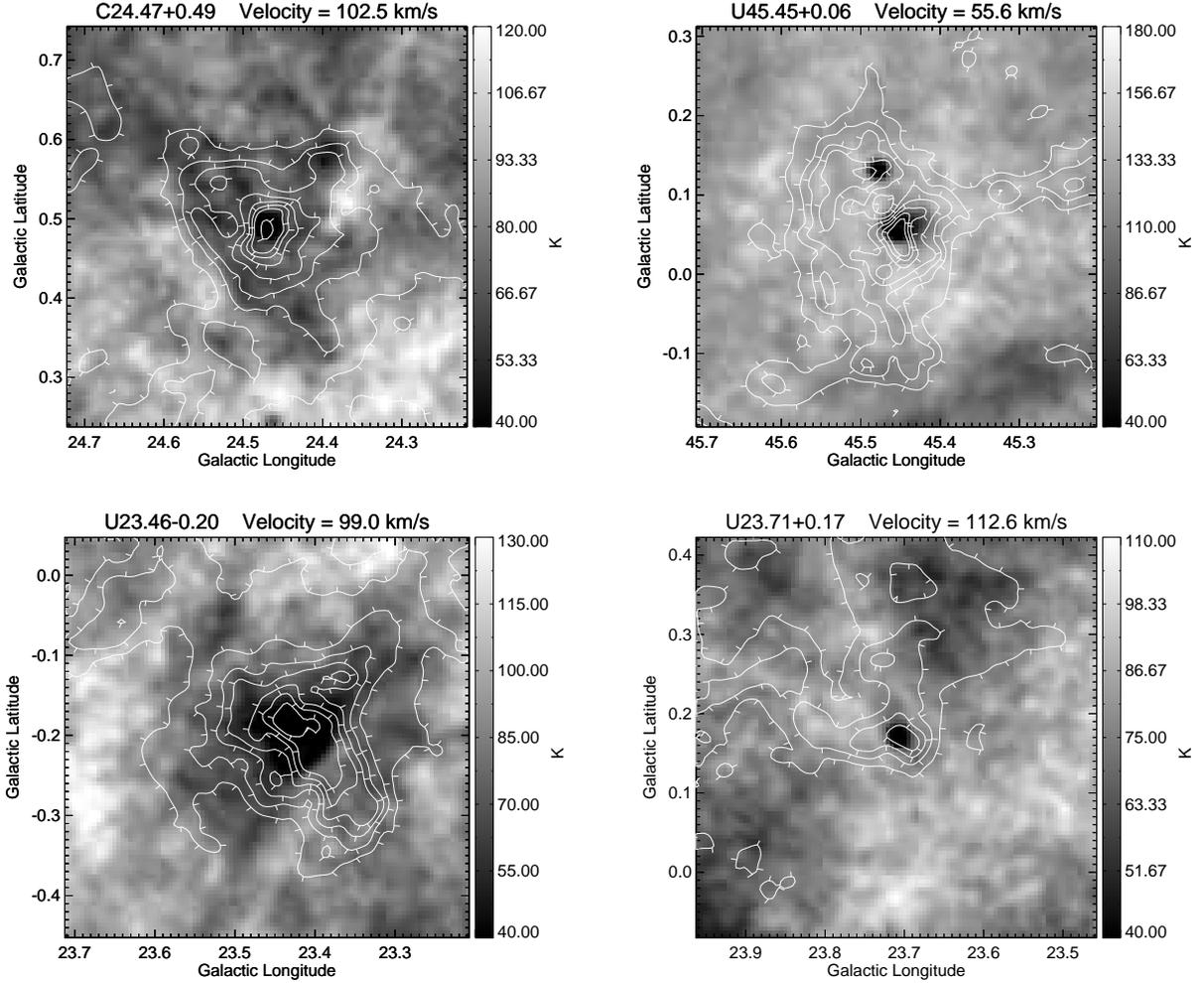}

\caption{Single channel VGPS images for the Figure \ref{fig:hisa} \hii\
regions at the velocity of peak \cor\ intensity used in the \hisa\
analysis.  The contours are of integrated \cor\ intensity, $W_{\rm
13CO}$.  Tick marks on the contours point towards decreasing values of
$W_{\rm 13CO}$.  Top row: example near (left) and far (right) sources
which we assign with high confidence (confidence parameter = A).  Bottom
row: example near (left) and far (right) sources which we assign with
low confidence (confidence parameter = B).  The two top images and bottom
right image show strong absorption from the \hii\ region radio continuum
at image center; these positions with strong radio continuum emission
were not used in the \hisa\ analysis.}

\label{fig:hisa_images}
\end{figure}

Background fluctuations and absorption from cold \hi\ gas not in
molecular clouds confuse the spectral analysis by creating features that
appear to be \hisa\ signals in the spectra.  Additionally, since we are
examining the larger molecular cloud surrounding a molecular clump,
there is an increased risk of blending molecular clouds at the near and
the far distances that share the same velocity.  We therefore require
the morphology of the absorption in the \hi\ images to match that of the
\cor\ gas.  We verify the results of the \hisa\ spectral analysis by
examining single channel \hi\ VGPS or SGPS images centered at the
velocity of peak \cor\ emission, with \cor\ integrated intensity
contours overlaid.  Figure \ref{fig:hisa_images} shows representative
images for the same sources who spectra were shown in Figure
\ref{fig:hisa}.  As is clear from Figure \ref{fig:hisa_images}, \hisa\
features may be difficult to determine from \hi\ images alone.  The
spectral analysis is often easier to interpret, and we mainly use the
images to aid in confusing cases.

We assign each source a confidence parameter to indicate the likelihood
that the \hisa\ feature is an artifact of noise or of background
fluctuations.  This qualitative parameter indicates our confidence that
the \hisa\ KDA resolution is correct.  For ``A'' sources at the near
distance, the \hi\ absorption signal is strong, unambiguous, and at the
same velocity as the \cor\ emission line. For ``A'' sources at the far
distance, the \hi\ at the velocity of the \cor\ emission line is smooth
and shows no fluctuations that could be interpreted as weak absorption.
The reliability of the KDA resolution for ``B'' quality sources is less
certain due to a fluctuating \hi\ spectrum or weak \cor\ emission.

\subsection{The KDA Resolutions}
We are able to resolve the KDA for 266 out of 291 sources (91\%).  This
statistic is a bit misleading, though, as the percentage of successful
KDA resolutions is $99\%$ for UC \hii\ regions, 97\% for compact \hii\
regions, and $72\%$ for diffuse \hii\ regions.  Diffuse \hii\ regions
are difficult to analyze mainly due to their weak radio continuum and
molecular emission.  Additionally, the \hiea\ analysis is more difficult
for \hii\ regions of larger angular extent, since finding ``off'' positions
that measure the same background as the ``on'' positions is challenging.

The \hiea\ method has a few limitations that decrease the number of
sources whose KDA can be resolved.  For lines of sight with multiple
\hii\ regions, the \hiea\ method is unable to distinguish which \hii\
region is causing the absorption.  Additionally, we assign \hii\
regions with RRL velocities within 10 \kms\ of the tangent point
velocity to the tangent point distance.  Excluding the sources with
multiple velocities and those whose RRL velocity is within 10 \kms\ of
the the tangent point velocity, our success rate for the \hiea\ method
is 72\%.  The \hiea\ method is better suited to strong, compact radio
sources, which also favors UC and compact \hii\ regions.  We were able
to successfully resolve the KDA towards 84\% of UC regions, 89\% of
compact regions, and 29\% of diffuse regions.  Our success rate is
42\% when all sources are included.

The \hisa\ method is more successful than the \hiea\ method at making
KDA determinations.  We were able to resolve the KDA for 87\% of {\it
all} \hii\ regions: 99\% of UC, 89\% of compact, and 64\% of diffuse
\hii\ regions.  Paper I found that UC and compact \hii\ regions have
similar, and strong, molecular emission properties.  This leads to a
higher success rate for these regions in the \hisa\ analysis.

Our KDA resolutions are approximately in the ratio of 2 to 1 for far
verses near distances.  There is about twice as much Galactic plane
area within the solar circle spanned by far KDA distances compared to
near distances over our longitude range.  Assuming a uniform surface
density of \hii\ regions, this ratio of 2 to 1 is what one would
expect from geometric effects alone.  The ratio of far to near sources
is highest for UC and diffuse \hii\ regions (2.2 and 3.8,
respectively), and significantly lower for the compact \hii\ regions
(1.6).  Because the diffuse regions are large in angular extent, one
may expect that this large angular size is due to proximity to the
Sun.  These results imply that diffuse \hii\ regions are truly large
and extended: their population is evenly spread throughout the inner
Galaxy.  The low ratio of far to near compact \hii\ regions implies
two possibilities: (1) compact \hii\ regions are truly more common
close to the Sun, or (2) UC and compact \hii\ regions are not distinct
classifications and many compact \hii\ regions would be classified as
ultra compact if placed at the far distance.

We believe that option (1) above is unlikely, and that the UC and compact
\hii\ regions are not distinct classes of object.  The UC \hii\ regions
in our sample were found using infrared color criteria from the IRAS
point source catalog.  As shown by \citet{conti04}, IRAS color criterion
are not unique to UC \hii\ regions; giant \hii\ regions (\hii\ regions
with $> 10^{50}$ Lyman continuum photons s$^{-1}$) occupy the same
infrared color-space as UC \hii\ regions.  The distinction between UC
and compact \hii\ regions is therefore likely one largely of angular size.

The ratio of far to near sources is similar for both KDA resolution
methods; the ratio for all three \hii\ region classifications is 1.9
for the \hiea\ method and 2.1 for the \hisa\ method.  As mentioned in
\S \ref{sec:resolve}, a null result in the \hiea\ method (no
absorption between the RRL velocity and the tangent point velocity)
implies the near distance while a null result in the \hisa\ method (no
absorption) implies the far distance.
That these two methods return a similar ratio of far verses near sources
implies that our results are not heavily influenced by a lack of
absorption.

With resolved distance ambiguities, we can transform the \hii\ region's
velocity into a distance using a rotation curve.  We use the
\citet{mcclure07} rotation curve because it is the most densely sampled
rotation curve extent, and includes data from both the first and the
fourth Galactic quadrants.  Table \ref{tab:2} summarizes the parameters
derived by the resolution of the distance ambiguity for our sample of
291 \hii\ regions.  Listed are the source name, together with the
parameters derived from the \hiea\ and \hisa\ analyses, the KDA resolved
distance from the Sun, $D_\sun$ (kpc), and the height above the Galactic
plane, $z$ (pc).  The derived \hiea\ parameters include the maximum
velocity of \hi\ absorption, $V_{\rm max}$, whether \hi\ absorption was
detected within $10 \kms$ of the RRL velocity, the near/far KDA
resolution, and the confidence parameter for this determination, CEA.
The derived \hisa\ parameters include the near/far KDA resolution, and
the confidence parameter for this determination, CSA.  For sources whose
\hiea\ and \hisa\ KDA resolutions disagree, the KDA resolution that we
adopt is marked with an asterisk.  The distances for sources that the
\hiea\ analysis located at the tangent point and for which the \hisa\
analysis was able to resolve the KDA are the \hisa\ distances.  In
\S\ref{sec:agree} we describe how we resolve these discrepancies.

\subsection{H\titleI\ Emission-Absorption verses H\titleI\ Self-Absorption}
\begin{figure}
\plotone{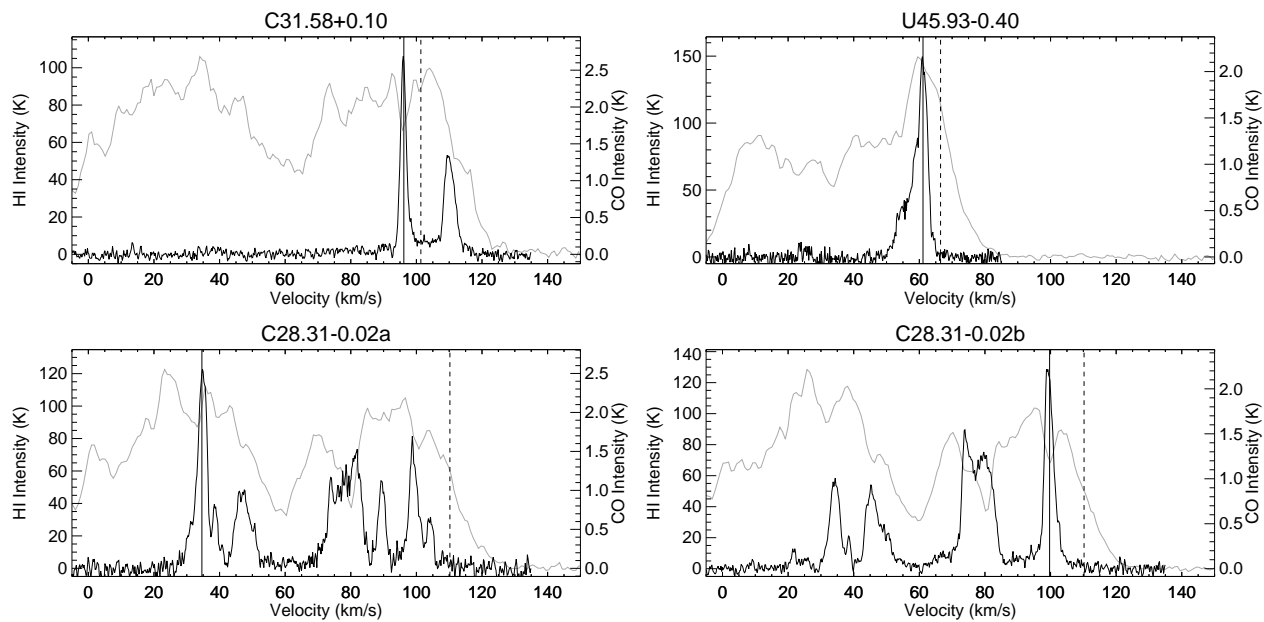}

\caption{Example \hisa\ analysis for four \hii\ regions whose KDA cannot
be resolved using the \hiea\ method.  The plot formats are those of
Figure \ref{fig:hisa}.  In the top row, the source velocities are very
near the tangent point velocities, and yet the \hisa\ analysis is able
to resolve the KDA; the top left source is at the near distance while
the top right source is at the far distance.  In the bottom row, two
\hii\ regions are located along the same line of sight, but the \hisa\
analysis is able to resolve the KDA for both sources.; C28.31$-$0.02a is
at the far distance while C28.31$-$0.02b is at the near distance.}

\label{fig:problems}
\end{figure}

The \hisa\ analysis can resolve the KDA for sources that cannot be
resolved using the \hiea\ method.  For the \hiea\ analysis we make no
attempt to resolve the KDA for sources within 10 \kms\ of the tangent
point velocity, as any determination would be unreliable.  Because
\hisa\ relies on background clouds at the same velocity, the reliability
of \hisa\ is independent of the RRL velocity.  The top row of Figure
\ref{fig:problems}, shows the \hisa\ spectra of C31.58$+$0.10 and U45.93$-$0.40,
sources with RRL velocities within 10 \kms\ of the tangent point
velocity.  For C31.58$+$0.10, there is significant absorption at the
same velocity as the associated CO gas, implying the near distance.
For U45.93$-$0.40, however, there is no absorption at the RRL
velocity, implying the far distance.

The \hisa\ method can also resolve the KDA for lines of sight with
multiple \hii\ regions.  If there are two \hii\ regions along a line of
sight, the \hiea\ analysis cannot determine which \hii\ region is
causing the absorption.  The bottom row of Figure \ref{fig:problems}
illustrates this point using the \hii\ regions C28.31$-$0.02a and
C28.31$-$0.02b.  This line of sight has two \hii\ regions, one at 35.8
\kms\ and one at 92.4 \kms\ \citep{lockman89}.  Figure
\ref{fig:problems} shows how \hisa\ can be used to resolve the KDA for
this source.  The \hii\ region at 35.8 \kms\ (C28.31$-$0.02a) does not
show absorption at the RRL velocity and must lie at the far distance.
The \hii\ region at 92.4 \kms\ (C28.31$-$0.02b) does show absorption at
the RRL velocity, and thus lies at the near distance.

The use of both methods gives a more robust \hii\ region KDA resolution.
The \hiea\ method solves the KDA for 6 sources that could not be
resolved using the \hisa\ method.  We are able to resolve the KDA for an
additional 138 sources using the \hisa\ method.  Of these 138 sources,
72 were placed at the tangent point in the \hiea\ analysis, 36 are lines
of sight with multiple \hii\ regions, and 30 did not show absorption
above the noise level in the \hiea\ analysis.

\subsection{Agreement Between H\titleI\,Self-Absorption and H\titleI\ Emission/Absorption \label{sec:agree}}
In general our two methods are in good agreement.  For sources not
assigned to the tangent point in the \hiea\ analysis, and for which
associated \cor\ was detected in Paper I, we find an 79\% agreement rate
between the two methods.  We estimate the robustness of each method by
comparing sources that have ``A'' confidence parameters in one method
against all sources in the other.  For our \hiea\ ``A'' sources, the
\hisa\ KDA determination agrees 84\% of the time.  The same analysis
using ``A'' \hisa\ sources shows an 97\% agreement with the \hiea\
analysis results.  Although the \hisa\ method is able to resolve the KDA
for more sources (cf. \S\ref{sec:discussion}), it appears to be less
robust than the \hiea\ method.  \citet{busfield06} found in single
pointing CO measurements towards 94 massive young stellar objects that
the \hisa\ method was $\sim 80\%$ accurate, in agreement with our results.

\begin{figure}
\plotone{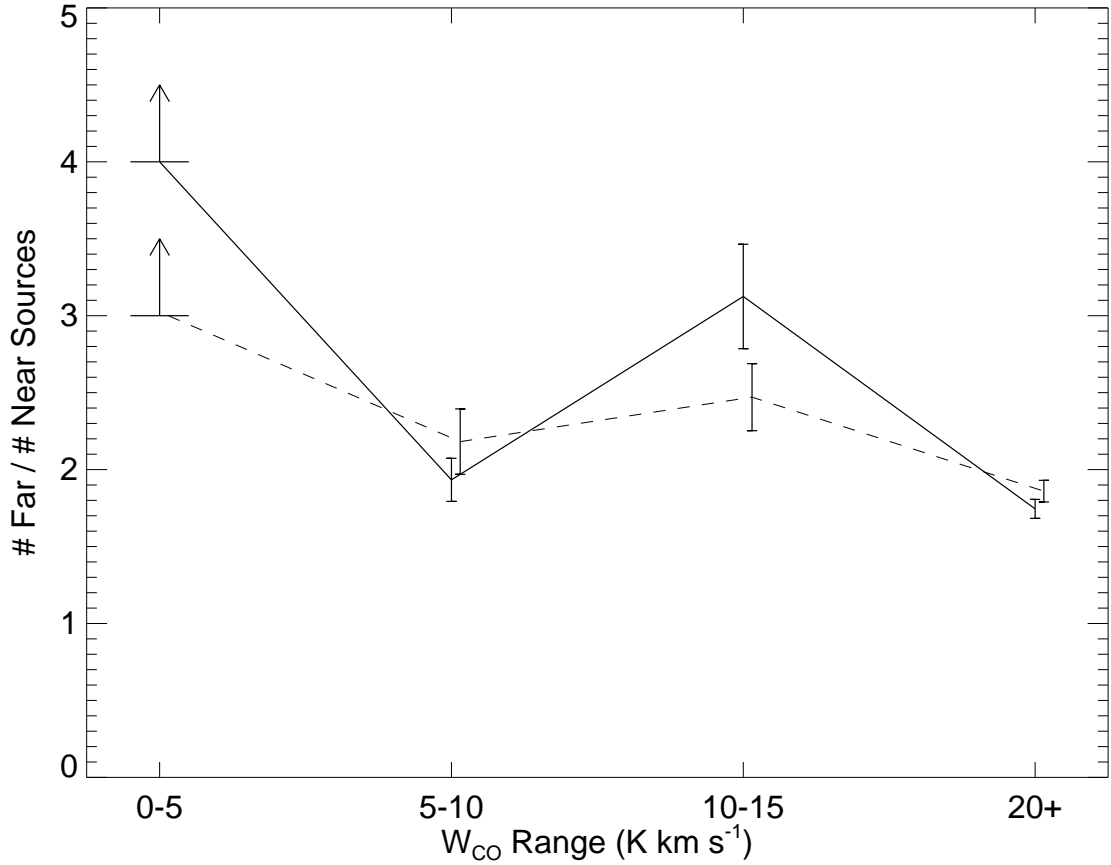}

\caption{The ratio of far to near sources plotted as a function
of the average \hii\ region integrated intensity, $W_{\rm 13CO}$.  The
solid line shows this relationship for a sample of \hii\ region/\cor\
cloud complexes defined by 75\% of the peak \cor\ value.  The
dashed line shows the relationship defined by 85\% of the peak \cor\
value.}

\label{fig:wco_vs_nf}
\end{figure}

There are many possible reasons why an \hiea\ and \hisa\ KDA resolution
may disagree for any given \hii\ region:
\begin{enumerate}

\item Molecular gas at the near distance may not produce \hisa\ due to
warming of the \cor\ gas.  This appears to be the case for
U23.96+0.15, U25.38$-$0.18, and U34.26+0.15.  For these regions, the
\hiea\ analysis is unambiguous and places the \hii\ region at the near
distance.  The molecular material is of a morphology and intensity that
a misassignment by Paper I is unlikely.  These are well known bright
\hii\ region complexes with high infrared luminosity.  The molecular
material surrounding these regions is probably heated and thus produces
a poor absorption signal.  All of these regions have associated \cor\
gas with higher than average excitation temperatures (see Paper I).  If
$T_s = 30 \K$, instead of 10~K (as we assumed in \S \ref{sec:hisa}), we
would need a $W_{\rm 13CO}$ value $\sim 30\%$ higher to produce the same
absorption signal.

\item Molecular gas associated with \hii\ regions may not produce \hisa\
because of insufficient column density of cold \hi.  In \S
\ref{sec:hisa}, we calculate that a $W_{\rm 13CO}$ value of $\sim
10$~K~\kms\ is required to produce measurable absorption ($W_{\rm
13CO}$ is proportional to column density).  Sources that do not
produce absorption are assigned to the far distance, possibly in
error.  If an insufficient column density of \cor\ is an issue, the
ratio of far to near sources should increase with decreasing values of
$W_{\rm 13CO}$.  We plot in Figure \ref{fig:wco_vs_nf} the ratio of
far to near sources verses the average \hii\ region $W_{\rm 13CO}$
value.  In this figure, the $W_{\rm 13CO}$ values are binned by 5 K
\kms\ intervals.  The ratio of far to near sources shown in the solid
line results from the 75\% threshold criterion, whereas the dashed
line results from the 85\% threshold criterion.  The $0-5$ K \kms\
bin does show a higher ratio of far to near determinations.  There are no near sources in either
the $75\%$ threshold bin nor the $85\%$ threshold bin; the ratio is
undefined.  There are, however, only 5 sources in the $75\%$ $0-5$ K
\kms\ bin and 4 sources in the $85\%$ bin.  The sources in
the 0$-$5 K \kms\ bin all have a confidence parameter value of ``B'' and
therefore their \hisa\ resolutions are less robust.  We conclude that
\hisa\ suffers no near/far bias if the average $W_{\rm 13CO}$ value is
at least 5 K \kms.  Therefore, \hisa\ can be caused even by relatively
low column densities of \cor.

\item There may be no \hisa\ signal because of a lack of warm background
\hi.  The observable result of such a situation would be a lack of
absorption at the velocity of \cor\ emission.


\item Cold \hi\ at the near distance not associated with molecular
clouds may cause a \hisa\ signal when the \hii\ region lies at the far
distance.  This appears to be the case for U42.11$-$0.44.  In this
situation, a \hii\ region at the far distance is incorrectly assigned
to the near distance in the \hisa\ analysis.
\hisa\ is frequently caused by \hi\ gas not inside molecular clouds
\citep[see ][]{peters83, bania84, peters87}.  \citet{gibson00} found
that most \hisa\ features in the 21cm \hi\ Canadian Galactic Plane
(outer Galaxy) Survey, a survey that is very similar to the VGPS and SGPS, have
no obvious \co\ counterpart.
We attempt to remove errors caused by cold \hi\ not in molecular clouds
by analyzing the VGPS and SGPS ({\it l, b\/}) images at the velocity of
the molecular gas.  We require the absorbing \hi\ gas to have a similar
spatial morphology as the \cor\ emission.

\item Velocity blending may confuse the \hisa\ analysis by combining two
unrelated clouds along the line of sight.  We used larger molecular
clouds in our analysis, instead of the molecular clumps identified in
Paper I.  The emission from a molecular cloud associated with an \hii\
region at the near distance may be blended with that of a molecular
cloud at the far distance, diminishing the \hisa\ signal.  Conversely, a
molecular cloud at the far distance may be confused with molecular
emission at the near distance that is associated with an \hii\ region,
thus causing a \hisa\ signal.  We inspected the ({\it l, b}) single
channel VGPS and SGPS images to look for the signature of velocity
crowding: clouds that go from absorption into emission.  This visual
signature, however, could also be caused by a variable background, and
therefore velocity crowding may remain a problem.
\item A lack of cold \hi\ between the \hi\ region and the tangent point velocities
may lead to inaccuracies in the \hiea\ analysis.  \citet{bania84}
estimate that there is a cold \hi\ cloud every 0.7 kpc, on average.  At
the low end of our longitude range, $l=15\arcdeg$, our criterion that
places sources at the tangent point if their RRL velocity is within 10
\kms\ of the tangent point velocity corresponds to about 0.7 kpc.  At
the high end of our longitude range, $l=56\arcdeg$, it corresponds to
about 2 kpc.  Therefore, we do not believe that a lack of \hi\ between
the source and the tangent point is a large problem.

\item One may mistake \hisa\ for \hiea.  The two methods do not
necessarily have unique spectral signatures, because in both cases the line width is
supersonic.  Visual inspection of the integrated intensity maps should
reduce errors from this confusion.  If the absorption is caused by the
\hii\ region's radio continuum, the morphology will match that of the
21cm continuum emission; if it is caused by the \hi\ inside molecular
clouds, it will match the \cor\ emission.

\item Fluctuations (from either the background or from noise) may result
in a false absorption signal in the \hiea\ analysis.  The error from
background and noise fluctuations should also be minimized from visual
inspection of the integrated intensity maps.  Background features that
may look like absorption features in the spectra will not match the 21cm
continuum emission when seen in the ({\it l, b}) maps.

\item Finally, the 20\% disagreement rate may be caused by inaccurate
\cor/\hii\ associations in Paper I.  For example, the molecular material
may lie at the near distance whereas the \hii\ region is at the far
distance.  It is difficult to estimate the occurrence of inaccurate
\cor/\hii\ associations because of the many possible errors just
discussed.  There are probably a very small number of mis-associations,
however, as the aforementioned errors can account for many of the
discrepancies between the results of the \hiea\ and \hisa\ methods.
\end{enumerate}


We believe the \hiea\ KDA analysis to be more robust than the \hisa\
analysis for \hii\ regions.  \hiea\ directly measures the absorption
from the free-free emission of an \hii\ region.  The \hiea\ KDA
resolutions are more robust because the presence of absorption at
velocities beyond the RRL velocity makes the KDA resolution unambiguous.
Also, the absorption caused by \hiea\ can be much stronger than that
caused by \hisa\ as the continuum brightness temperature of \hii\
regions at 21cm can be very high.  Therefore, a smaller column density
of \hi\ may produce absorption in the \hiea\ method compared to what is
required in the in \hisa\ method.  Our \hisa\ method is less direct, and
uses molecular gas as a proxy for emission from the \hii\ region.  The
use of \cor\ as a proxy has many potential uncertainties, as enumerated
above.  The KDA resolutions made using the \hisa\ method remain slightly
ambiguous even if absorption is present because there are many ways to
create absorption at the same velocity as \cor\ emission.

For sources with a disagreement between the two methods, our final KDA
determination is that found by the method with a confidence parameter
value of ``A''.  If both methods have the same confidence parameter
value, we visually examine the spectra and images for evidence that one
method has a more robust result.  In general, we choose the distance
determination of the \hiea\ method.

\subsection{Agreement with Other Distance Determinations}
Our KDA resolutions agree with most previously published distance
determinations.  We agree with $\sim 75\%$ of the determinations made
previously using ${\rm H_2CO}$ absorption, and agree with $\sim 95\%$ of
determinations made previously using \hiea.  The adopted tangent point
velocity depends on the adopted rotation curve, and therefore is not
uniform from study to study.  In our comparisons with previously
published studies below, we disregard \hii\ regions located at the
tangent point in either study to create a fair comparison.  The sources
for which our KDA resolution disagrees with a previously published KDA
resolution are summarized in Table \ref{tab:3}, which lists the source
name, our determination, the other author's determination, and the
reference.

\subsubsection{${\rm H_2CO}$ Absorption Distances}
Our \hii\ region sample shares 50 non-tangent point nebulae with
\citet{downes80}.  We have 14 disagreements in KDA resolution, and one 
source which has multiple velocity components, U25.72+0.05.  The
second velocity for U25.72+0.05 was unknown to \citet{downes80}.
Seven of these sources \citet{downes80} assign to the near distance
while we assign the far distance.  These seven discrepancies can be
explained if ${\rm H_2CO}$ is not detectable between the \hii\ region
and the tangent point.  The remaining seven discrepancies are
difficult to explain.  The rather high RMS noise of the VGPS and SGPS
\hi\ data, however, may be a factor.


We have 10 non-tangent point sources in common with \citet{wink82}, and
disagree on three determinations, for: U19.61$-$0.24, U43.18$-$0.52, and U43.89$-$0.78.  All
three sources \citet{wink82} locate at the near distance, while we
locate at the far distance.

We have 9 non-tangent point sources in common with \citet{araya02},
and disagree with three of their determinations.  We disagree on
U35.57+0.07, U35.58$-$0.03, and U50.32+0.68, all of which
\citet{araya02} locate at the near distance.  U35.58$-$0.03 was
reobserved by the same group
\citep{watson03} and their newer determination agrees with ours.
\citet{kuchar94} agrees with our determination for U35.57+0.07, which we
locate at the far distance with high confidence in both the \hisa\ and
\hisa\ analyses.  U50.32+0.68 is a very weak continuum source, but does
appear to show true absorption at 68 \kms\ in our \hiea\ analysis,
implying the far distance.

We have 23 non-tangent point sources in common with \citet{watson03}.
We disagree with five of their determinations for these 23 sources.  We
disagree on U50.32+0.68, which was mentioned previously.  We are
confident in our determination for U34.09+0.44 and U35.67-0.04, as the
determinations in both \hiea\ and \hisa\ agree with high confidence.  The \hisa\ analysis for U35.02+0.35 shows strong \hi\
absorption near the velocity of \cor\ emission, although the
velocities are very slightly different.  We were unable to perform the
\hiea\ analysis on U34.40+0.23 since it is a very weak radio continuum
source, but the \hisa\ analysis implies the far distance.  For
U34.40+0.23, this is particularly strange since it appears to be
associated with an infrared dark cloud.  As infrared dark clouds are
seen in silhouette against the infrared background, they most likely
lie at the near distance.  This could be a case where \hisa\ is
produced within the cloud itself, or the \hii\ region is not
associated with the infrared dark cloud.

We have 20 non-tangent point sources in common with \citet{sewilo04},
and disagree on three determinations.  We disagree with their
determination for U23.27+0.08, C23.24$-$0.24, and U24.50$-$0.04.  The
\hii\ region U23.27+0.08 is quite weak and we were unable to perform
the \hiea\ analysis.  Using the \hisa\ analysis we assign the near distance,
although with low confidence.  For C23.24$-$0.24, we assign the near
distance in both analyses, while \citet{sewilo04} assign the far
distance based on the detection of a molecular cloud in absorption at
95.6 \kms.
We detect absorption at 119 \kms in our \hiea\ analysis for U24.50$-$0.04, implying the far
distance.  \citet{sewilo04} argue for the near distance.   This \hii\ region is so
close to the tangent point, however, that the KDA resolution has little
impact on the assigned distance.

\subsubsection{{\rm H{\small I}} Emission/Absorption Distances}
We share 32 non-tangent point determinations with
\citet{kuchar94} and disagree on  U44.26+0.10 and C51.36+0.00.
The \hii\ region U44.26+0.10 is too close to the tangent point for the
\hiea\ analysis, but the \hisa\ analysis places it at the far
distance.  Our \hisa\ analysis also shows that C51.36+0.00 is at the
far distance.  There is an absorption line in the \hisa\ analysis near
the velocity of peak \cor\ emission, but this line is caused by a
large \hisa\ feature that is not associated with \cor.

We have 36 non-tangent point sources in common with \citet{kolpak03}.
We disagree on the KDA resolution for U23.71+0.17, C25.41$-$0.25, and
C30.78$-$0.03.  For U23.71+0.17, our two method disagree.  The \hiea\
method favors the near distance, since the absorption lines are quite
strong and appear only up to the RRL velocity.  Unfortunately,
C25.41$-$0.25 is too weak a continuum source for us to perform the
\hiea\ analysis.  The \hisa\ analysis of this source implies the far
distance with low confidence.  \citet{downes80} also assign the far
distance to C25.41$-$0.25.  For C30.78$-$0.03, we detect absorption at
97 \kms, which is within 10 \kms\ of the 91.6 \kms\ RRL velocity.
\citet{kolpak03} detect absorption at 122 \kms\ that is not present in
our analysis.  In the \hisa\ analysis, there is a strong absorption
feature at the same velocity and with the same line width as a \cor\
feature, implying the near distance.

We share six non-tangent point sources in common with \citet{fish03},
and agree on all except for U28.20-0.05.  This source lies near the
tangent point, but our \hisa\ analysis shows no \hi\ absorption at the
velocity of \cor\ emission, which implies the far distance.


\subsection{Catalog of H\titleII\ Region Properties}
As was done in Paper I, we provide our results online in an \hii\ region
catalog\footnote[2]{http://www.bu.edu/iar/hii\_regions}.  This website
now contains all the molecular line data from Paper I, as well as all
data from Tables \ref{tab:1} and \ref{tab:2}, the spectra from which we
made the distance determinations in both the \hisa\ and \hiea\ analyses,
and images we used to assist our KDA resolution.  This website should
become a valuable tool for the study of \hii\ regions.

\section{DISCUSSION \label{sec:discussion}}
\begin{figure}

\plotone{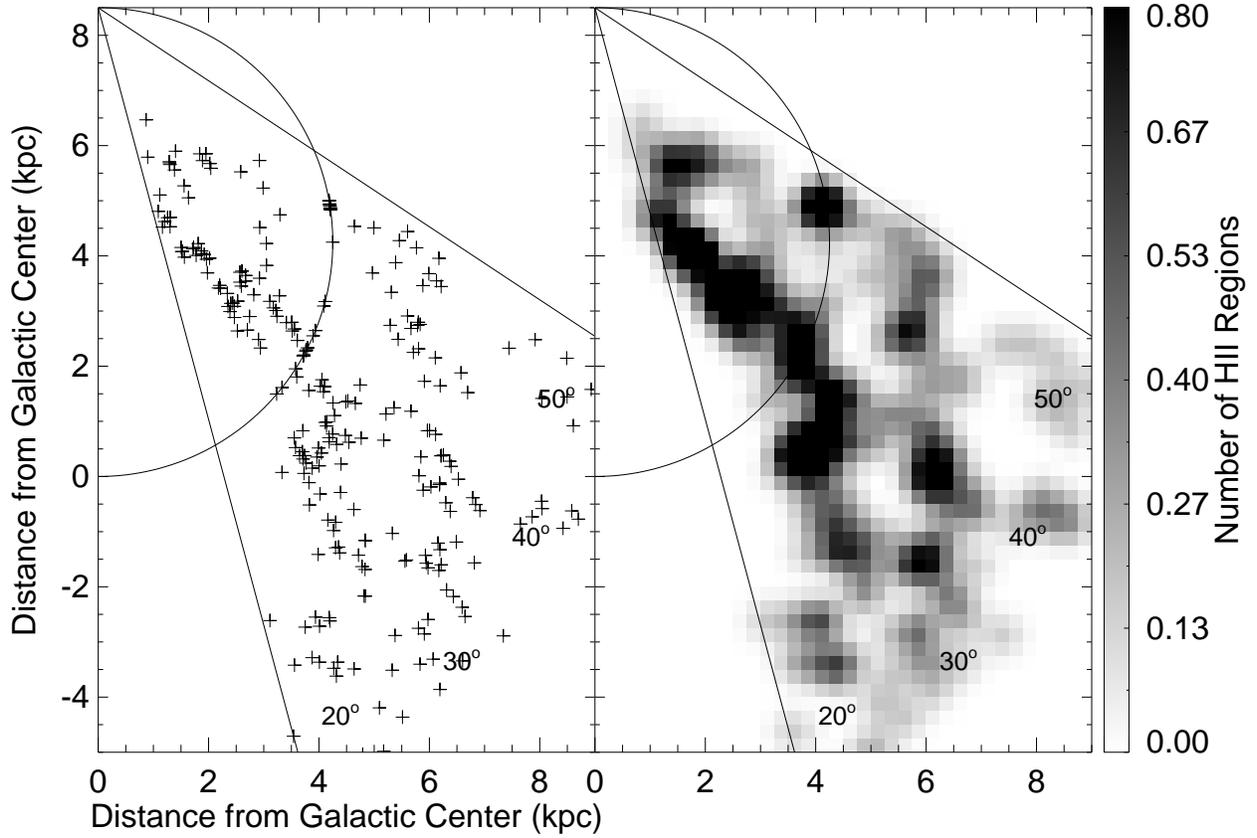}

\caption{Galactic Distribution of \hii\ regions.  The straight lines
show the longitude range of our sample, $15\arcdeg-56 \arcdeg$.  The
half-circle indicates the tangent point distance, and is the locus of
subcentral points.  The left panel plots shows the positions for all
regions with KDA resolved distances.  In the right panel, these
positions are binned into $0.25 \times 0.25$ \kpc\ pixels, and the
resultant image smoothed with a $5 \times 5$ pixel Gaussian
filter.}

\label{fig:faceon}
\end{figure}

\begin{figure}

\plotone{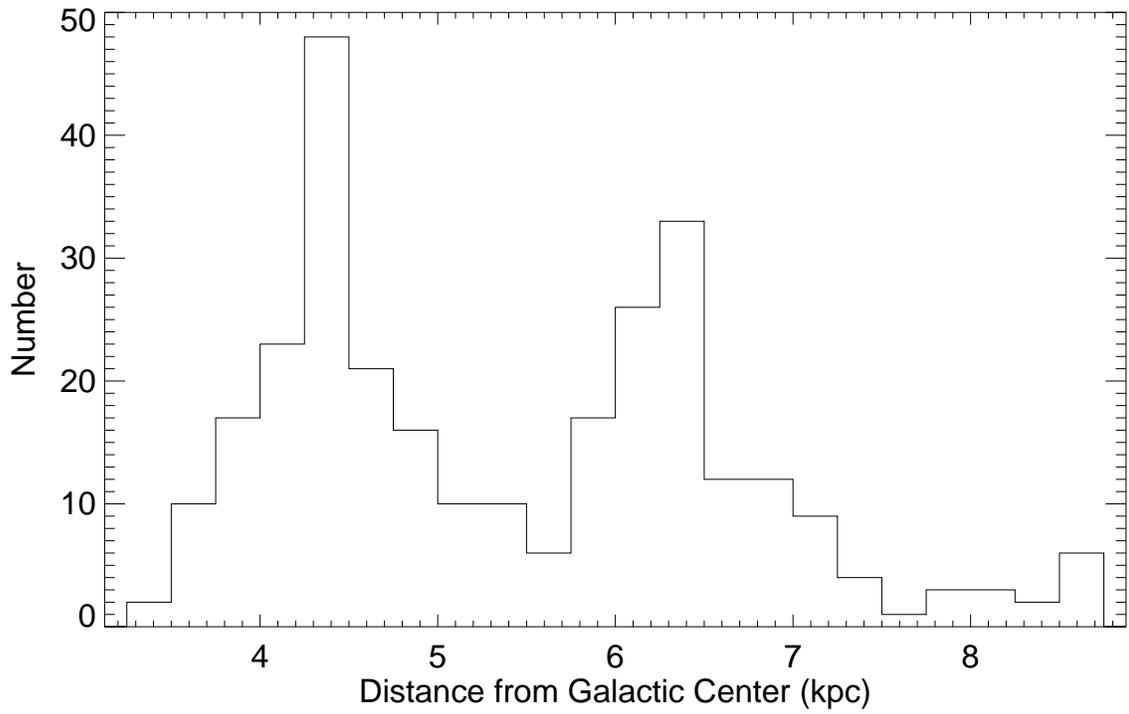}

\caption{Distribution of Galactocentric radii for our \hii\ region
sample.  The distribution is shown using 0.25 kpc bins.  The peaks at
4.5 and 6 kpc in this distribution correspond to the radii of the
circular arc features seen in Figure \ref{fig:faceon}.}

\label{fig:rgal}
\end{figure}

\begin{figure}

\plotone{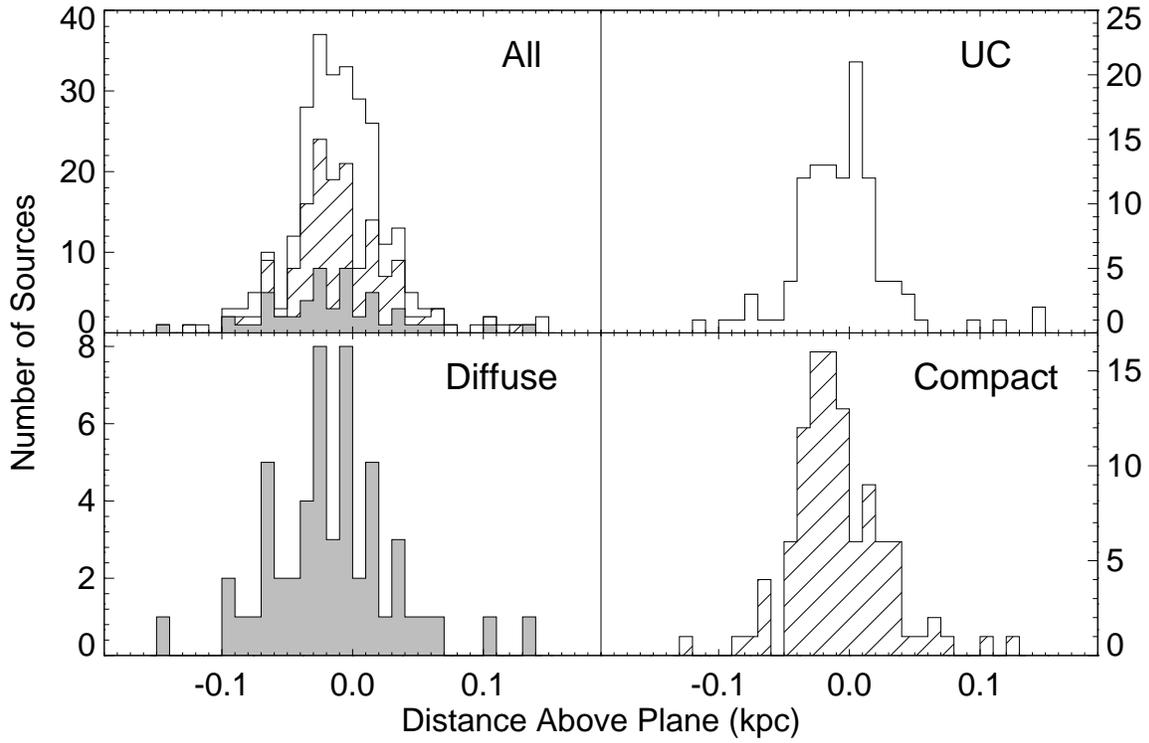}

\caption{Distribution of height above the plane for the sources with
resolved distance ambiguities.  Ultra compact sources are shown open, compact
sources are shown hatched, and diffuse sources are shown in gray.  The
top left panel is the stacked histogram of all three categories of \hii\
regions such that the outer line shows the entire distribution. The
other three panels show the distribution for the three categories of
\hii\ regions.}

\label{fig:height}
\end{figure}

The distribution of the 266 \hii\ regions with resolved distance
ambiguities projected onto the Galactic plane is shown in Figure
\ref{fig:faceon}.  The left panel of Figure \ref{fig:faceon} shows a
scatter plot of the data.  In the right panel, the data are binned into
$0.25 \times 0.25$ kpc pixels, and then smoothed with a 5 pixel Gaussian
filter.
Figure \ref{fig:faceon} appears to show some hints of Galactic
structure.  There are two circular arc segments centered at the Galactic
Center, near where the Scutum arm (at a Galactocentric radius of
$\sim 4.5$ kpc) and the Sagittarius arm (at a Galactocentric radius of
$\sim 6$ kpc) are thought to be located.  The most striking features of
this plot though are where the \hii\ regions are {\it not} located,
namely at a Galactocentric radius of $\sim 5$ kpc and within 3.5 kpc of
the Galactic Center.  We believe the paucity of \hii\ regions within 3.5
kpc of the Galactic center is further evidence of a Galactic bar of
half-length 4 kpc, as described by \citet{benjamin05}.  This region is
very well sampled, with almost 100 \hii\ regions within $l = 25\arcdeg$.
These results are supported by the histogram of Galactocentric radii
shown in Figure \ref{fig:rgal}.  An \hii\ region's Galactocentric radius
is a function only of the rotation curve, and not of any KDA distance
resolution.  Figure \ref{fig:rgal} shows concentrations of \hii\ regions
at Galactocentric radii of 4.5 and 6 kpc.  There are large streaming
motions found at these radii \citep{mcclure07}.
The distribution of height above the Galactic plane for \hii\ regions
with resolved distance ambiguities is shown in Figure \ref{fig:height}.
The top left panel of Figure \ref{fig:height} is a stacked histogram;
the top curve in this panel shows the entire distribution and the
shadings represent the proportion of this total made up by the
categories of \hii\ regions.  Clockwise from top right are the
individual histograms for the UC, compact, and diffuse nebulae.  The
entire distribution appears to be Gaussian, with a FWHM of 68~pc (the
standard deviation of the z height, $\sigma_z$, is 42 pc), centered at
$-11$~pc.
This result is slightly lower than that found by other authors, who
generally find an offset of $\sim 20$~pc \citep[see ][and references
therein]{reed06}.  While we are only sampling $|b| < 1\arcdeg$ in this
study, the number of \hii\ regions outside the latitude range is small
and should not affect our results significantly if included in our
sample.

One may expect that UC regions, being younger, would have had less time
to drift from their natal environments, and thus would have a narrow
distribution of heights above the Galactic plane.  Our results show that
UC, compact, and diffuse \hii\ regions in fact share similar
distributions, with $\sigma_z$ values near 40 pc.  A Gaussian fit to
each distribution shows some segregation, with FWHM values of 60 pc, 64
pc, and 96 pc for UC, compact and diffuse regions, respectively.  A
Kolmagorov-Smirnov (K-S) test, however, reveals that these differences
are not significant.  The K-S test assesses the likelihood that two
samples are drawn from the same parent distribution.  We find a 35\%
probability that the UC and compact \hii\ region scale heights are drawn
from the same parent distribution, a 15\% probability for the compact
and diffuse \hii\ region scale heights, and a 9\% probability for
compact and diffuse \hii\ region scale heights.

Values of $\sigma_z$ near 40 pc have been found by many authors.  In
their study of UC and compact \hii\ regions, \citet{paladini04} showed
that for the sources with the most secure distance determinations,
$\sigma_z \simeq 40$ pc.  In a study of UC \hii\ region candidates,
\citet{bronfman00} estimate $\sigma_z \simeq 30$ pc.  \citet{giveon05}
estimate in their sample of UC \hii\ region candidates that $\sigma_z
\simeq 30$ pc as well.  Neither the \citet{bronfman00} nor the
\citet{giveon05} study makes distance determinations for their \hii\
regions, and therefore these scale height estimates may be uncertain.

\section{SUMMARY}
We resolve the kinematic distance ambiguity (KDA) for 266 inner Galaxy
\hii\ regions out of a sample of 291 using the 21cm \hi\ VLA Galactic
Plane Survey and the \cor\ BU-FCRAO Galactic Ring Survey.  Our sample of
\hii\ regions is divided into three subsets: ultra compact (UC),
compact, and diffuse nebulae.  We use two methods to resolve the
distance ambiguity for each \hii\ region: \hi\ Emission/Absorption
(\hiea) and \hi\ self-absorption (\hisa).  There is an 79\% agreement
rate between the two methods.  We find that the \hiea\ method is more
robust than the \hisa\ method for \hii\ regions, but the \hisa\ method
is able to resolve the distance ambiguity for almost twice the number of
sources.  We estimate the robustness of the \hiea\ and \hisa\ methods
as implemented here at 97\% and 84\%, respectively.  Using \hisa\ we can
resolve the distance ambiguity for lines of sight with two \hii\ regions
at different velocities, and for \hii\ regions near the tangent point.
We find that the \hisa\ signal can be caused by modest column densities
of \cor\ ($W_{\rm 13CO}$ values down to $\sim 5$ K \kms) and therefore
the utility of this method is not limited to large, dense molecular
clouds.  We have greater success for both methods with UC and compact
\hii\ regions, as their strong radio continuum emission and association
with molecular material make the distance determinations unambiguous in
most cases.

Our sample of \hii\ regions is approximately in the ratio 2 to 1 for far
verses near KDA resolutions.  The ratio of far to near UC and diffuse
\hii\ regions is 2.2 to 1 and 3.8 to 1, respectively. For compact \hii\ regions, the
ratio is 1.6 to 1.  This implies that compact \hii\ regions are not a
physically defined classification, but rather that many compact \hii\
regions are so classified because of their proximity to the Sun.  The
diffuse \hii\ regions' large angular sizes are not due to proximity to
the Sun.

Our KDA resolutions agree with most previously published results.  Our
results agree with $\sim 95$\% of the KDA resolutions found previously
using the \hiea\ method, and with $\sim 75$\% of those found using ${\rm
H_2CO}$ absorption.  Some of the disagreements with the results of ${\rm
H_2CO}$ absorption studies can be explained by the relatively low
line of sight filling factor of ${\rm H_2CO}$.

Our sample of \hii\ regions appears to trace aspects of Galactic
structure such as circular arc segments at Galactocentric radii of 4.5
and 6 kpc, and a lack of \hii\ regions within 3.5 of the Galactic
center.  There is a paucity of \hii\ regions at a Galactocentric
radius of 5 kpc.  These features are largely independent of any distance
determination.  The scale height for all three \hii\ region
classifications is around 40 pc.



\acknowledgments Here we use molecular line data from the Boston
University-FCRAO Galactic Ring Survey (GRS). The GRS is a joint project
of Boston University and the Five College Radio Astronomy Observatory,
funded by the National Science Foundation under grants AST-9800334,
AST-0098562, and AST-0100793.  We also use \hi\ data from the VLA
Galactic Plane Survey (VGPS).  The VGPS is supported by a grant from the
Natural Sciences and Engineering Research Council of Canada and from the
National Science Foundation. 
\nraoblurb


\clearpage

\begin{deluxetable}{lccccccc}
\tabletypesize{\scriptsize}
\tablecaption{\hii\ Region Kinematic Parameters}
\tablewidth{0pt}
\tablehead{

\colhead{Source} & 
\colhead{l} & 
\colhead{b} & 
\colhead{$V_{\rm LSR}$} & 
\colhead{$R_{\rm GC}$} &
\colhead{$D_{\rm near}$} & 
\colhead{$D_{\rm far}$} & 
\colhead{$D_{\rm TP}$} \\

\colhead{} &
\colhead{($\arcdeg$)} &
\colhead{($\arcdeg$)} &
\colhead{($\kms$)} &
\colhead{(kpc)} &
\colhead{(kpc)} &
\colhead{(kpc)} &
\colhead{(kpc)} \\
}

\startdata

U23.87$-$0.12  &  23.87  &  $-$0.12  & \phn 73.8  &  4.6  &    4.7  &  10.8  &  7.8  \\
C23.91+0.07a  &  23.91  &  $+$0.07  & \phn 32.8  &  6.4  &    2.4  &  13.1  &  7.8  \\
C23.91+0.07b  &  23.91  &  $+$0.07  & 103.4  &  3.8  &    6.1  &  \phn 9.4  &  7.8  \\
U23.96+0.15  &  23.96  &  $+$0.15  & \phn 78.9  &  4.4  &    5.0  &  10.6  &  7.8  \\
C24.00$-$0.10  &  24.00  &  $-$0.10  & \phn 75.8  &  4.5  &    4.8  &  10.7  &  7.8  \\
C24.13$-$0.07  &  24.13  &  $-$0.07  & \phn 86.9  &  4.2  &    5.3  &  10.2  &  7.8  \\
C24.14+0.12  &  24.14  &  $+$0.12  & 114.5  &  3.6  &    6.8  &  \phn 8.7  &  7.8  \\
D24.14+0.43  &  24.14  &  $+$0.43  & \phn 98.4  &  4.0  &    5.9  &  \phn 9.6  &  7.8  \\
C24.19+0.20  &  24.19  &  $+$0.20  & 111.9  &  3.7  &    6.6  &  \phn 8.9  &  7.8  \\
C24.22$-$0.05  &  24.22  &  $-$0.05  & \phn 82.0  &  4.4  &    5.1  &  10.4  &  7.8  \\

\enddata
\label{tab:1}
\tablecomments{Table 1 is published in its entirety in the electronic edition of the {\it Astrophysical Journal}. 
A portion is shown here for guidance regarding its form and content.}
\end{deluxetable}

\begin{deluxetable}{lccccccccc}
\tabletypesize{\scriptsize}
\tablecaption{\hii\ Region Kinematic Distances}
\tablewidth{0pt}
\tablehead{

\colhead{} & 
\multicolumn{4}{c}{\hiea} &
\colhead{} &
\multicolumn{2}{c}{\hisa} &
\colhead{} & 
\colhead{} \\ \cline{2-5} \cline{7-8}

\colhead{Source} & 
\colhead{$V_{\rm max}$} & 
\colhead{Abs?} & 
\colhead{N/F} & 
\colhead{CEA} & 
\colhead{} & 
\colhead{N/F} & 
\colhead{CSA} & 
\colhead{$D_{\sun}$} & 
\colhead{$z$} \\

\colhead{} &
\colhead{($\kms$)} &
\colhead{} &
\colhead{} &
\colhead{} &
\colhead{} &
\colhead{} &
\colhead{} &
\colhead{(kpc)} &
\colhead{(pc)} \\
}

\startdata

U23.87$-$0.12  &  100  &  Y  &  F\phantom{*}  &  A  & &  F\phantom{*}  &  A  & 10.8  &  \phn $-22.6$  \\
C23.91+0.07a  &  \nodata  &  \nodata  &  \nodata\phantom{*}  &  \nodata  & &  F\phantom{*}  &  B  & 13.1  &  \phn \phs 16.1  \\
C23.91+0.07b  &  \nodata  &  \nodata  &  \nodata\phantom{*}  &  \nodata  & &  F\phantom{*}  &  B  & \phn 9.4  &  \phn \phs 11.5  \\
U23.96+0.15  &  \phn 82  &  Y  &  N*  &  A  & &  F\phantom{*}  &  B  & \phn 5.0  &  \phn \phs 13.0  \\
C24.00$-$0.10  &  102  &  Y  &  F\phantom{*}  &  A  & &  F\phantom{*}  &  A  & 10.7  &  \phn $-18.7$  \\
C24.13$-$0.07  &  109  &  Y  &  F\phantom{*}  &  A  & &  F\phantom{*}  &  A  & 10.2  &  \phn $-12.4$  \\
C24.14+0.12  &  \nodata  &  \nodata  &  T\phantom{*}  &  \nodata  & &  F\phantom{*}  &  A  & \phn 8.7  &  \phn \phs 18.3  \\
D24.14+0.43  &  \nodata  &  \nodata  &  \nodata\phantom{*}  &  \nodata  & &  N\phantom{*}  &  A  & \phn 5.9  &  \phn \phs 44.1  \\
C24.19+0.20  &  \nodata  &  \nodata  &  T\phantom{*}  &  \nodata  & &  F\phantom{*}  &  A  & \phn 8.9  &  \phn \phs 31.1  \\
C24.22$-$0.05  &  119  &  Y  &  F\phantom{*}  &  A  & &  F\phantom{*}  &  B  & 10.4  &  \phn \phn $-9.1$  \\

\enddata
\label{tab:2}
\tablecomments{Table 2 is published in its entirety in the electronic edition of the {\it Astrophysical Journal}. 
A portion is shown here for guidance regarding its form and content.}

\end{deluxetable}

\begin{deluxetable}{lccl}
\tabletypesize{\scriptsize}
\tablecaption{Distance Discrepancies}
\tablewidth{0pt}
\tablehead{

\colhead{Source} & 
\colhead{This Work} & 
\colhead{Past Work} & 
\colhead{Reference} \\
}

\startdata

C19.61-0.13 & F & N & 1 \\
U19.61-0.24 & F & N & 1, 2 \\
U19.68-0.13 & F & N & 1 \\
U20.08-0.14 & F & N & 1 \\
C22.76-0.49 & N & F & 1 \\
C22.95-0.32 & N & F & 1 \\
C22.98-0.36 & N & F & 1 \\
C23.24-0.24 & N & F & 5 \\
U23.27+0.08 & N & F & 5 \\
C23.54-0.04 & N & F & 1 \\
U23.71+0.17 & N & F & 7 \\
U23.96+0.15 & N & F & 1 \\
C25.41-0.25 & F & N & 7 \\
U24.50-0.04 & F & N & 5 \\
U25.38-0.18 & N & F & 1 \\
U28.20-0.05 & F & N & 8 \\
C30.78-0.03 & N & F & 7 \\
U31.40-0.26 & F & N & 1 \\
C23.24-0.24 & N & F & 1 \\
U34.09+0.44 & F & N & 4 \\
U34.40+0.23 & F & N & 4 \\
U35.02+0.35 & N & F & 4 \\
U35.57+0.07 & F & N & 3 \\
U35.58-0.03 & F & N & 1, 3 \\
U35.67-0.04 & F & N & 1, 4 \\
U43.18-0.52 & F & N & 2 \\
U43.89-0.78 & F & N & 2 \\
U44.26+0.10 & F & N & 6 \\
U50.32+0.68 & F & N & 3, 4 \\
C51.36+0.00 & F & N & 6 \\

\enddata

\tablenotetext{1}{\citet{downes80}}
\tablenotetext{2}{\citet{wink82}}
\tablenotetext{3}{\citet{araya02}}
\tablenotetext{4}{\citet{watson03}}
\tablenotetext{5}{\citet{sewilo04}}
\tablenotetext{6}{\citet{kuchar94}}
\tablenotetext{7}{\citet{kolpak03}}
\tablenotetext{8}{\citet{fish03}}

\label{tab:3}
\end{deluxetable}

\clearpage

\end{document}